\documentclass[11pt]{article}
\usepackage{amsfonts,amscd,amsmath, amssymb,graphicx}
\newcommand{\oh}{\frac{1}{2}}
\def\z{z_{\text{\tiny 0}}}
\def\ep{\text{e}}
\def\zt{z_{\text{\tiny T}}}
\def\T{T_c}
\def\lt{l_{\text{\tiny T}}}
\def\g{\mathfrak{g}}
\def\oh{\frac{1}{2}}

\title{The Spatial String Tension in the Deconfined Phase of  $SU(N)$ Gauge Theory and Gauge/String Duality}
\author{Oleg Andreev \thanks{Also at Landau Institute for Theoretical Physics, Moscow. }\\\\
{\it Technische Universit\"at M\"unchen, }\\
{\it Cluster of Excellence for Fundamental Physics,} \\
{\it Boltzmannstrasse 2, 85748 Garching, Germany}
}
\date{}

\begin{document}

\maketitle
\begin{abstract}
The spatial string tension of a $SU(N)$ gauge theory without quarks is calculated using gauge/string duality for 
$1.2\,\T\leq T\leq 3\,\T$. The result is remarkably consistent with the available lattice data for $N=2,\,3$. 
Some evidence is found that to leading order the large distance physics of a string can be described by a five-dimensional 
geometry without a need for an internal space.
 \end{abstract}

\vspace{-13 cm}
\begin{flushright}

{\small SPAG-A1/07}\\
\end{flushright}

\vspace{13 cm}


\section{Introduction}
\renewcommand{\theequation}{1.\arabic{equation}}
\setcounter{equation}{0}

It is known that a pure $SU(N)$ gauge theory at high temperature undergoes a phase transition. Although at the phase transition 
point basic thermodynamical observables show drastic qualitative change, there are some whose structure does not change qualitatively at 
$\T$. For example, this is the case for the pseudo-potential extracted from spatial Wilson loops. It is confining for all temperatures \cite{deconf}. 
This is taken as an indication that certain confining properties survive in the high temperature phase. Indeed, thermal perturbation theory works 
well only for high temperatures, while non-perturbative effects make it difficult to do computations near the transition point. In addition, the 
results obtained at RHIC for $\T\leq T\leq2\,\T$ indicate that the matter created in heavy ion collisions is strongly coupled \cite{RHIC}. Thus, 
there is a need for new approaches to strongly coupled gauge theories. 

Until recently, the lattice formulation even struggling with limitations and systematic errors was the main computational tool to deal with 
strongly coupled gauge theories. The situation changed drastically with the invention of the AdS/CFT correspondence that resumed interest 
in finding a string description of strong interactions. In this Letter we will explore the temperature dependence of the spatial string tension by 
using gauge/string duality. The temperature range of interest is $1.2\,\T\leq T\leq 3\,\T$. The lower limit is chosen to keep the system out of 
the critical regime. The upper limit is determined by consistency. One of the motivations for this investigation is to subject the 
proposal of \cite{andreev} to further test. Another is to get the explicit expression for the tension which is in general difficult to calculate in 
strongly coupled gauge theories. 

Following \cite{andreev}, we take the following ansatz for the 10-dimensional background geometry which turned out to be
quite successful in describing the thermodynamics of pure gauge theories in the temperature range we are considering\footnote{Equation 
\eqref{metric} is the ansatz. This time we do not know equations that provide such a solution. So, we follow "the inverse problem'': first, we
suggest a solution, then we look for its phenomenological relevance.}

\begin{equation}\label{metric}
ds^2
=h\,\frac{R^2}{z^2}
\left(fdt^2+d\vec x^2+\frac{1}{f}dz^2\right)+
\frac{1}{h}\,R^2 d\Omega_{\text{\tiny 5}}^2
\,, \qquad
f=1-\frac{z^4}{\zt^4}
\,,\qquad
h=\ep^{\frac{1}{2}cz^2}
\,,
\end{equation}
where $\zt=1/\pi T$. $c$ is a positive and $N$-dependent parameter to be fixed from the critical temperature.\footnote{For
$SU(3)$, it may also be determined from the $\rho$ meson trajectory. The discrepancy in the value of the critical temperature is of
order $10$  \% \cite{andreev}. } In fact, the metric is a deformed product of the Euclidean $\text{AdS}_5$ black hole and
the 5-dimensional sphere $S^5$ considered as an internal space. The deformation is due to a $z$-dependent factor $h$. Such a deformation is 
crucial for breaking conformal invariance of the original supergravity solution and introducing $\Lambda_{\text{\tiny QCD}}$. We also take a
constant dilaton.

\section{Calculating the Spatial String Tension}
\renewcommand{\theequation}{2.\arabic{equation}}
\setcounter{equation}{0}

Given the background metric, we can calculate expectation values of Wilson loops by using the proposal of \cite{malda}.
The loops obeying an area law provide string tensions. Our goal is therefore to study spatial Wilson loops.\footnote{The literature on the Wilson 
loops within the AdS/CFT correspondence is very vast. For an earlier discussion of the spatial loops, see, e.g., \cite{list} and references therein.}

To this end, we consider a rectangular loop ${\cal C}$ along two spatial directions $(x,y)$ on the boundary $(z=0)$ of 10-dimensional space.
As usual, we take one direction to be large, say $Y\rightarrow\infty$. The quark and antiquark are set at $x_i=\pm\tfrac{r}{2}$, respectively.
 In the internal space the string lies along the great circle of the sphere. So, we can take $\Theta$ as an angle along the circle and set the
particles at $\Theta_i=\pm\tfrac{\theta}{2}$.

Now we will use the Nambu-Goto action with the background metric \eqref{metric}. We choose the world-sheet coordinates as $\xi_1=x$ and
$\xi_2=y$. The action is then

\begin{equation}\label{NG}
S=\frac{\g}{2\pi}Y
\int ^{\frac{r}{2}}_{-\frac{r}{2}} dx\,\frac{h}{z^2}
\sqrt{1+\frac{1}{f} \,z'^2+\frac{z^2}{h^2}\,\Theta'^2}
\,,
\end{equation}
where $\g=\tfrac{R^2}{\alpha'}$. A prime denotes a derivative with respect to $x$. The boundary conditions are
given by

\begin{equation}\label{boundary}
z\left(\pm\tfrac{r}{2}\right)=0\,,\qquad
\Theta\left(\pm\tfrac{r}{2}\right)=\pm \frac{\theta}{2}
\,.
\end{equation}

From the symmetries of the problem we see that the Euler-Lagrange equations have first integrals

\begin{equation}\label{f-integrals}
\frac{h}{z^2}\Bigl(1+\frac{1}{f} \,z'^2+\frac{z^2}{h^2}\,\Theta'^2\Bigr)^{-\oh}=\text{constant}
\,,\qquad
\frac{z^2}{h^2}\Theta'=\text{constant}
\,.
\end{equation}
The integration constants can be expressed via the values of $z$ and $\Theta'$ at $x=0$. A useful observation is that $z$ reaches its maximum
value $\z$ at $x=0$ and, as a result, $z'\vert_{x=0}=0$.

By virtue of \eqref{f-integrals}, the integrals over $\left[-\tfrac{r}{2},\tfrac{r}{2}\right]$ of $dx$ and $d\Theta$ are equal to

\begin{align}
r&=2\sqrt{\frac{l}{c}}\int_0^1 dv\, v^2 \exp\Bigl\lbrace \oh l(1-v^2)\Bigr\rbrace
\Bigl(1-\frac{l^2}{\lt^2}v^4\Bigr)^{-\oh}
\Bigl(1+k-kv^2-v^4\exp\bigl\lbrace l(1-v^2)\bigr\rbrace\Bigl)^{-\oh}
\,,\label{r}\\
\theta&=2\sqrt{k}\int_0^1 dv\exp\Bigl\lbrace \oh l v^2\Bigr\rbrace
\Bigl(1-\frac{l^2}{\lt^2}v^4\Bigr)^{-\oh}
\Bigl(1+k-kv^2-v^4\exp\bigl\lbrace l(1-v^2)\bigr\rbrace\Bigl)^{-\oh}
\,,\label{theta}
\end{align}
where for convenience we have introduced new dimensionless parameters $v=\frac{z}{\z}$, $l=c\z^2$,
$\lt=c\zt^2$, and  $k=\frac{z^2}{h^2}\Theta'^2\big\vert_{x=0}$.\footnote{Note that $l\geq 0$ and $k\geq 0$.}

At this point a comment is in order. A simple analysis shows that the integrals in \eqref{r} and \eqref{theta} are real for
the parameters subject to

\begin{equation}\label{1wall}
\lt>l
\end{equation}
and
\begin{equation}\label{2wall}
k>
\begin{cases}
l-2 & \text{if $\,2<l<2+\sqrt{2}$}\,,\\
2 (1+\sqrt{2})l^{-1}\exp\lbrace l-2-\sqrt{2}\rbrace
& \text{if $\,2+\sqrt{2}<l$}\,.
\end{cases}
\end{equation}
The first constraint is nothing else but a statement that the string may not go behind the horizon $(z=\zt)$, as should be for
the black hole geometry. This gives the first wall. The second says that at zero temperature the string is also
prevented from getting deeper into $z$ direction.\footnote{Note that for $k=0$ it reduces to the constraint $l<2$ described in \cite{az1}.} This
gives rise to the second wall. Both the walls provide the area law for the spatial Wilson loops. Now the question is which one is dominant for the
temperature range of interest? From the pont of view of \cite{az2}, where $k=0$, the second wall dominates. As we will see,
this is the case for $k\not=0$ too.

Now we will compute the energy (pseudo-potential) of the configuration. First, we reduce the integral over $x$ in \eqref{NG} to that over $z$. 
This is done by using the first integrals \eqref{f-integrals}. Since the integral is divergent at $z=0$ due to the factor $z^{-2}$ in the metric,  in 
the process we regularize the integral over $z$ by imposing a cutoff $\epsilon$. Then we replace $z$ with $v$ as in \eqref{r} and \eqref{theta}.
Finally, the regularized expression takes the form

\begin{equation}\label{energy}
E_{\text{\tiny R}}=
\frac{\g}{\pi}\sqrt{\frac{c}{l}}
\int_{\tfrac{\epsilon}{z_0}}^1 \frac{dv}{ v^2}
\exp\Bigl\lbrace \oh l v^2\Bigr\rbrace
\Bigl(1-\frac{l^2}{\lt^2}v^4\Bigr)^{-\oh}
\Bigl(1-\frac{k}{k+1}v^2-\frac{1}{k+1}v^4\exp\bigl\lbrace l(1-v^2)\bigr\rbrace\Bigl)^{-\oh}
\,.
\end{equation}
Its $\epsilon$-expansion is given by
\begin{equation*}\label{energy1}
E_{\text{\tiny R}}=\frac{\g}{\pi\epsilon}+E+O(\epsilon)
\,.
\end{equation*}
Subtracting the divergence (quark masses), we obtain a finite result
\begin{equation}\label{energy2}
E=\frac{\g}{\pi}\sqrt{\frac{c}{l}}
\int_0^1 \frac{dv}{ v^2}
\biggl[\exp\Bigl\lbrace \oh l v^2\Bigr\rbrace
\Bigl(1-\frac{l^2}{\lt^2}v^4\Bigr)^{-\oh}
\Bigl(1-\frac{k}{k+1}v^2-\frac{1}{k+1}v^4\exp\bigl\lbrace l(1-v^2)\bigr\rbrace\Bigl)^{-\oh}
-1-v^2\biggr]
\,.
\end{equation}
Similarly as $r$ and $\theta$, the pseudo-potential is real only for the parameters subject to the constraints \eqref{1wall}
and \eqref{2wall}.

\subsection{Numerical Analysis}

The pseudo-potential in question is written in parametric form given by Eqs.\eqref{r}, \eqref{theta}, and \eqref{energy2}. It is unclear to
us how to eliminate the parameters and find $E$ as a function of $r$ and $\theta$. We can, however, gain some important
insights from numerical calculations.

We start by noting that there is one more constraint on the allowed values of the parameters. It is simply
\begin{equation}\label{pi-constraint}
\theta\leq\pi
\end{equation}
but it is far from being trivial in terms of $l$ and $k$. An important point is that the constraints
\eqref{1wall} and \eqref{2wall} are satisfied for all the values of the parameters subject to \eqref{pi-constraint}.

In Fig.1 we have plotted the curve defined by $\theta (l,k)=\pi$
for several different values of $\lt$. The allowed values of the parameters lie between the $k$-axis and the curves.
%
\begin{figure}[ht]
\begin{center}
\includegraphics[width=4.5cm]{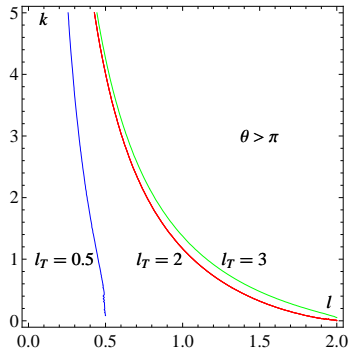}
\caption{\small{Typical graphs of $\theta (l,k)=\pi$ on the $(l,k)$ plane.}}
\end{center}
\end{figure}
Note that at $c=0$ or, equivalently $l=0$, $k$ can take any positive value. In this case $\theta$ is an increasing function of $k$ such that
$\theta=0$ at $k=0$ and $\theta\rightarrow\pi$ as $k$ goes to infinity.

Having determined the allowed region for the parameters, we are now ready to study $r$ as a function of two variables. To this end,
we look for level curves in the allowed region. A summary of our numerical results is shown in Fig.2.
\begin{figure}[htp]
\centering
\includegraphics[width=4.5cm]{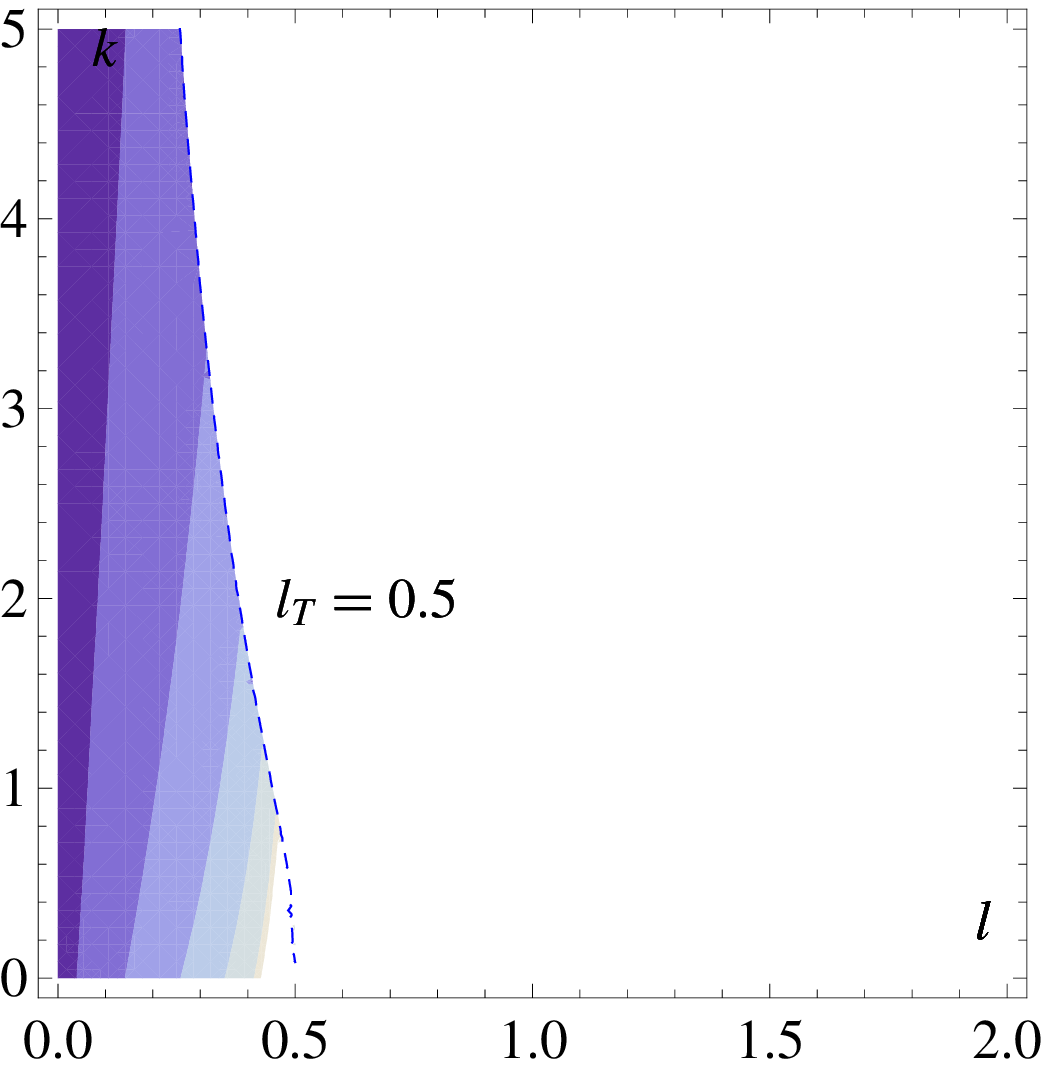}
\hfill
\includegraphics[width=4.5cm]{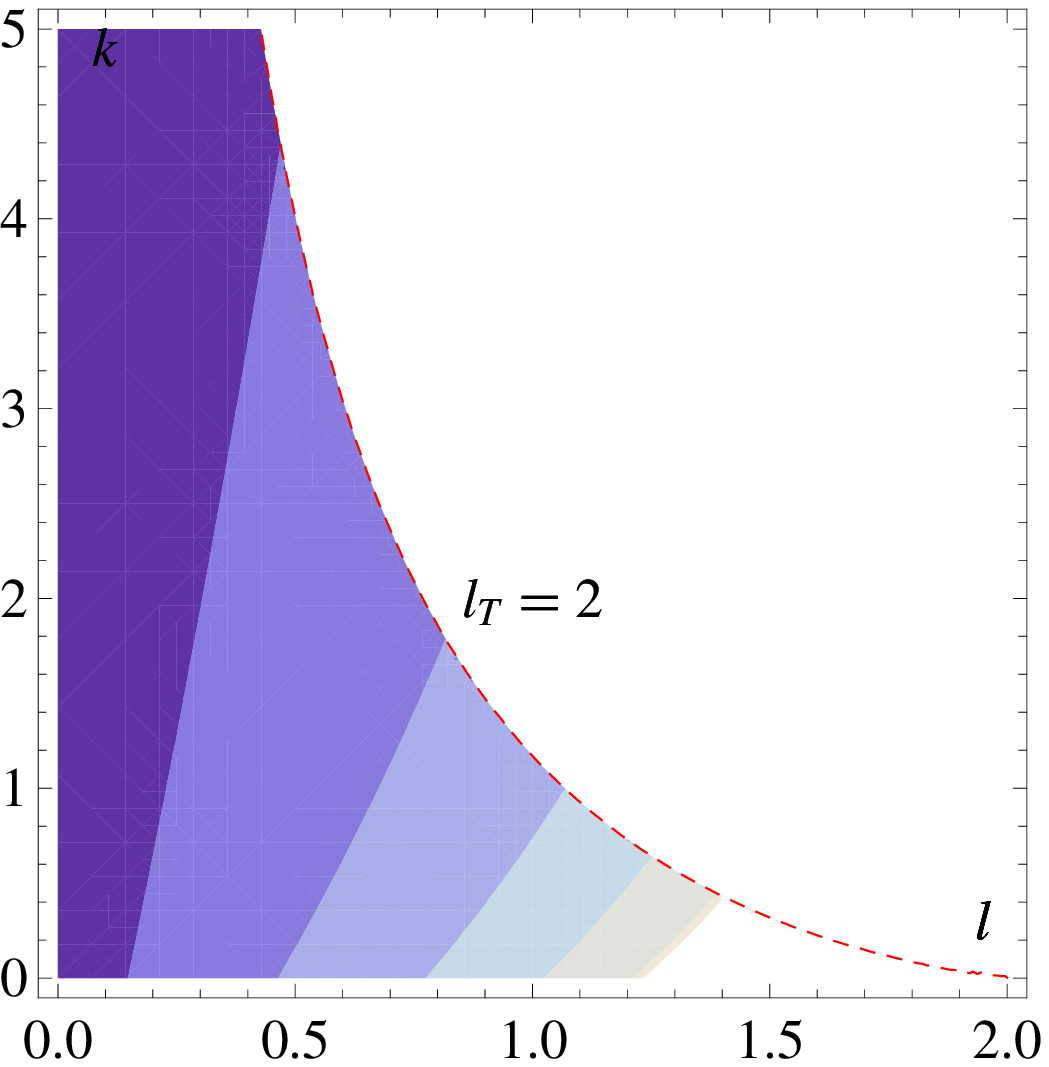}
\hfill
\includegraphics[width=4.5cm]{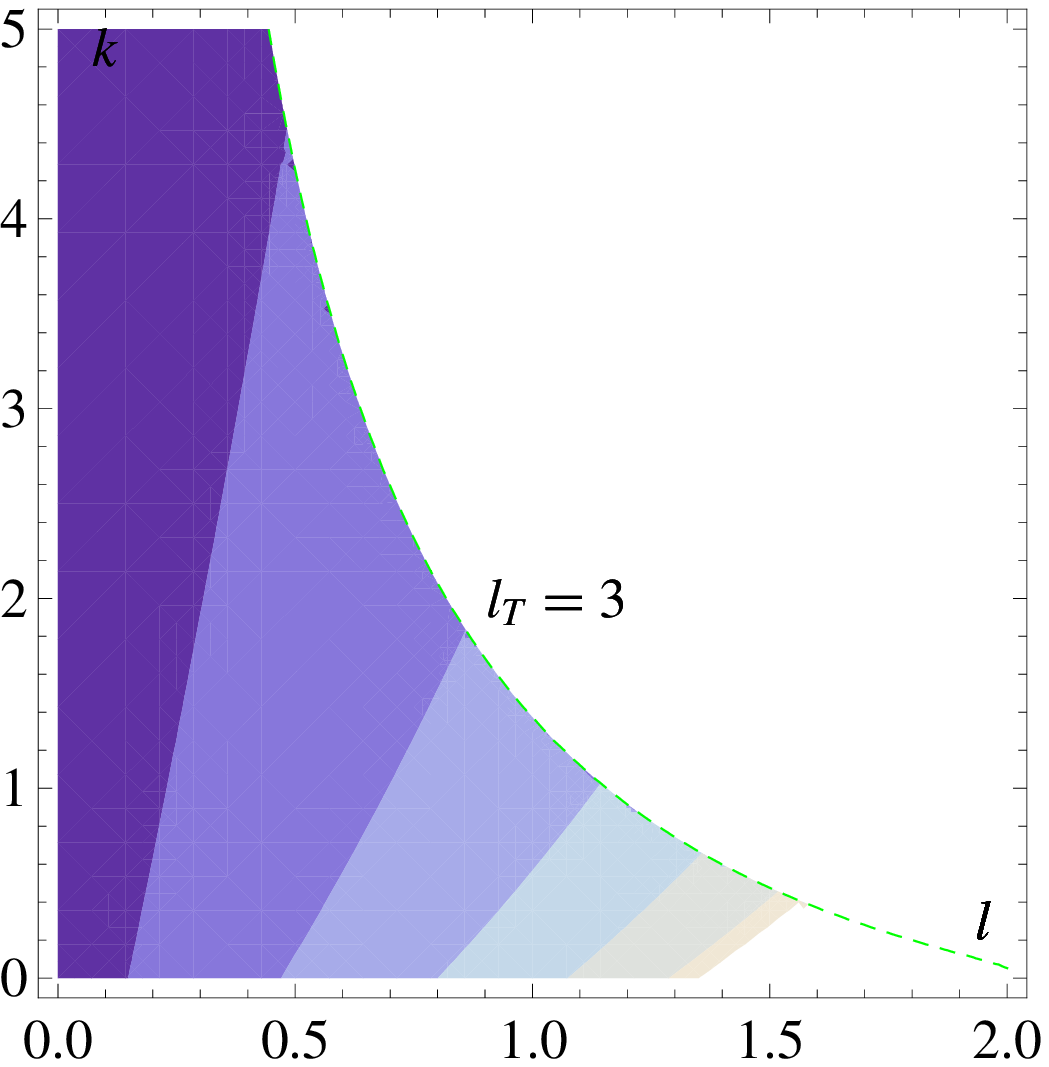}
\\
 \caption{\small{Level curves for $r=r(l,k)$. Larger values are shown lighter. The dashed lines show $\theta (l,k)=\pi$.}}
\end{figure}
The first observation is that $r$ takes large values only in the right corner of the region. Evidently, $k$ is small near the tip lying on 
the $l$-axis. An important point is that the tip is located at $l=2$ if $\lt\geq 2$ and at $l=\lt$ if $\lt\leq 2$. Since $k$ takes small values, it 
is not surprising that such a behavior pattern is similar to what was found in \cite{az2} for $k=0$. According to this paper, there are three 
possibilities: 1) $\lt >2$. In this case, the large distance physics of the string is determined by \eqref{2wall}. The phase is interpreted as the low 
temperature phase or the confined phase. 2) $\lt <2$. The large distance physics is now determined by the near horizon geometry of the metric 
as the wall \eqref{1wall} terminates the string. This phase is interpreted as the high temperature phase or the deconfined phase. 3) $\lt =2$. This 
is a transition point. In terms of $\T$ and $c$, the condition $\lt=2$ is equivalent to

\begin{equation}\label{Tc}
\T=\frac{1}{\pi}\sqrt{\frac{c}{2}}
\,.
\end{equation}

For our purposes, we follow this pattern. So, we consider $\lt >2$ as the high temperature phase and take \eqref{Tc} as a definition of
the critical temperature.

Finally, we present the result for the pseudo-potential at large distances. As seen from Fig.3, it is linear and
\begin{figure}[ht]
\begin{center}
\includegraphics[width=6.5cm]{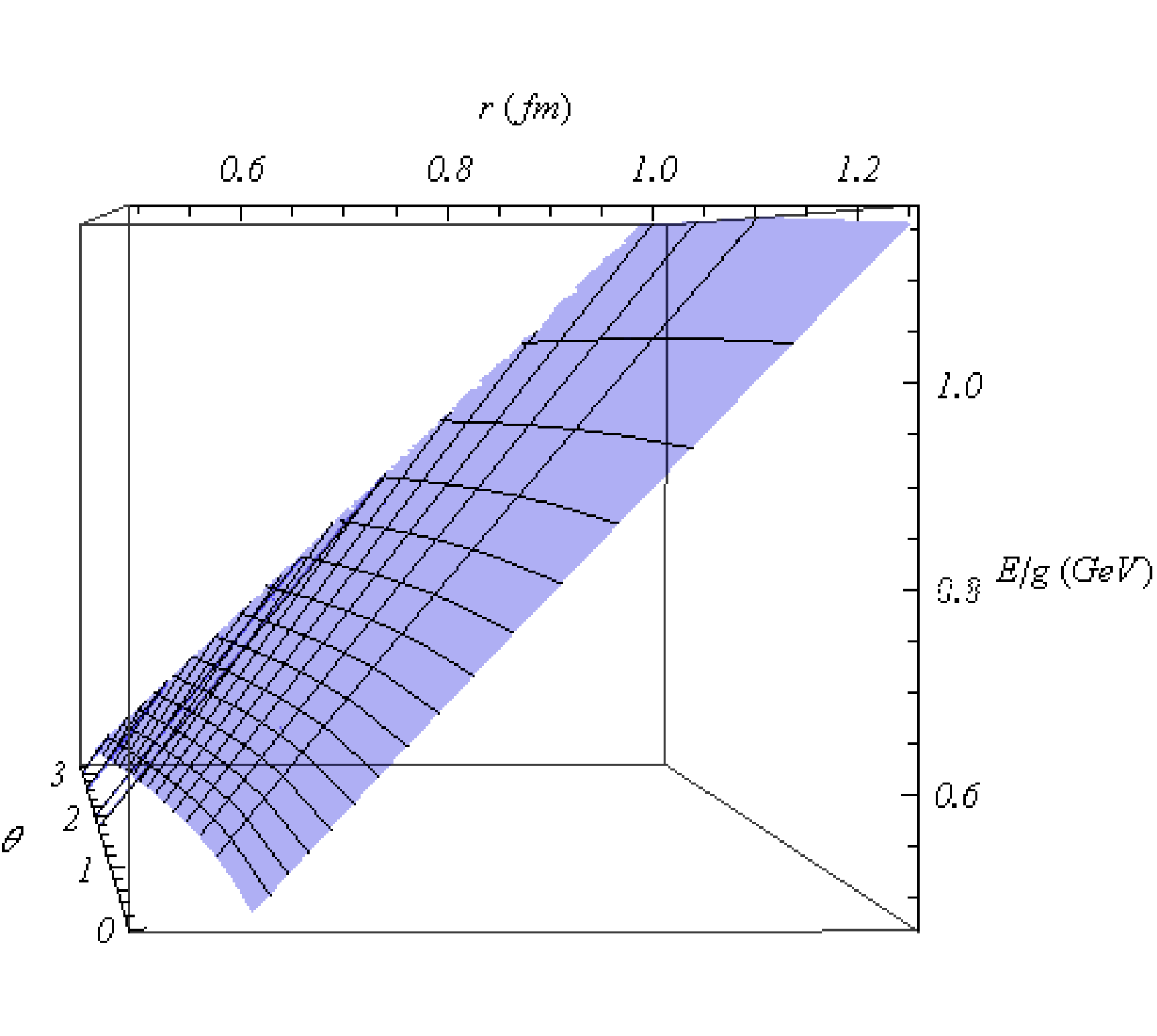}
\caption{\small{$E/\g$ as a function of $r$ and $\theta$. Here $c=0.9\,\text{GeV}^2$ and $\lt=\sqrt{2}$.}}
\end{center}
\end{figure}
the slope is almost independent of the angle $\theta$ between the positions of the quark and the antiquark in the internal space. So 
an optimistic view is to interpret the large distance physics of the string in terms of a five-dimensional geometry. In the case of interest it is 
given by the deformed Euclidean $\text{AdS}_5$ black hole. At this point a piece of evidence we can offer is that for $SU(2)$ the spatial string 
tension computed in \cite{az2} is remarkably consistent with the lattice results of \cite{bali}. Fortunately, we will see more evidence in the 
next sections.

\subsection{Analytical Calculations}
So we have gained the necessary knowledge. Now let us move on to the next step and compute the spatial string tension analytically. 

According to the numerical analysis, $r$ may become large only if $k$ is small. Expanding the right-hand sides of  \eqref{r}, \eqref{theta}, 
and \eqref{energy2} in $k$, to the leading order, we get  

\begin{align}
r&=2\sqrt{\frac{l}{c}}\int_0^1 dv\, v^2 \exp\Bigl\lbrace \oh l(1-v^2)\Bigr\rbrace
\Bigl(1-\frac{l^2}{\lt^2}v^4\Bigr)^{-\oh}
\Bigl(1-v^4\exp\bigl\lbrace l(1-v^2)\bigr\rbrace\Bigl)^{-\oh}
\,,\label{r-large}\\
\theta&=2\sqrt{k}\int_0^1 dv\exp\Bigl\lbrace \oh l v^2\Bigr\rbrace
\Bigl(1-\frac{l^2}{\lt^2}v^4\Bigr)^{-\oh}
\Bigl(1-v^4\exp\bigl\lbrace l(1-v^2)\bigr\rbrace\Bigl)^{-\oh}
\,,\label{theta-large}\\
E&=\frac{\g}{\pi}\sqrt{\frac{c}{l}}
\int_0^1 \frac{dv}{ v^2}
\biggl[\exp\Bigl\lbrace \oh l v^2\Bigr\rbrace
\Bigl(1-\frac{l^2}{\lt^2}v^4\Bigr)^{-\oh}
\Bigl(1-v^4\exp\bigl\lbrace l(1-v^2)\bigr\rbrace\Bigl)^{-\oh}
-1-v^2\biggr]
\,.\label{energy-large}
\end{align}
A nice thing is that the expressions \eqref{r-large} and \eqref{energy-large} being independent of $k$ are the same as those appeared in \cite{az2}. 
Therefore, we can borrow the results of this paper to discuss the long distance behavior of $E$. 

According to the analysis of \cite{az2}, $r$ becomes large in a region near $l=\lt$ where the integrals are dominated by the upper limits. After 
performing the integrals and eliminating the parameter $l$, one finds that at long distances the pseudo-potential is linear and the spatial string 
tension is given by 

\begin{equation}\label{stension}
\sigma_s=\sigma\,\frac{T^2}{\T^2}\, 
\exp\biggl\lbrace\frac{\T^2}{T^2} -1\biggr\rbrace
\,,
\end{equation}
where $\sigma=\frac{\g\,\ep}{4\pi}c$. Note that $\sigma$ is the physical string tension at zero temperature \cite{az1}.

However, this is not the whole story because of the remaining equation \eqref{theta-large}. A short calculation shows that for $l\sim\lt$ and 
$v\sim 1$ the right hand side of \eqref{theta-large} takes the form similar to that of \eqref{r-large}. This allows us to express $k$ in terms of $r$ 
and $\theta$. We get 

\begin{equation}\label{k}
k=\frac{1}{\pi^2 T^2}
\,\frac{\theta^2}{r^2}\,
\exp\biggl\lbrace -\frac{c}{\pi^2 T^2}\biggr\rbrace
\,.
\end{equation}
It follows from \eqref{k} that for finite $T$ and $\theta$, the parameter $k$ goes to $ 0$ as $r\rightarrow\infty$. Thus, our calculation is 
self-consistent. 

We conclude this subsection by making a couple of comments:

\noindent (i) The higher $k$-terms in the Taylor series for $r(k)$ and $E(k)$ result in the 
$\frac{\theta}{r}$-corrections to the pseudo-potential. So, the leading linear term is independent of the angle $\theta$ that
allows one to deal with it in framework of 5-dimensional gravity.\footnote{Note that in the 
case of $\alpha'$-corrected geometry a similar observation was made in \cite{dorn-otto}. } On the other hand, subleading 
corrections do depend on $\theta$. The latter implies that one has to involve an original 10-dimensional background to describe them. These 
arguments explain the large distance behavior of the pseudo-potential we observed in numerical calculations.

\noindent (ii) We should stress that the compactness of the internal space turns out to be crucial for the success of 5-dimensional gravity in 
the problem.

\subsection{Comparison with Lattice and Predictions}

It is of great interest to compare the temperature dependence of the spatial string tension \eqref{stension} with other results for the high 
temperature phase of $SU(N)$ gauge theory. In Fig.4 a comparison is shown between the square root of the spatial string tension from 
lattice simulations and the analogous quantity calculated analytically. 
%
\begin{figure}[ht]
\begin{center}
\includegraphics[width=6cm]{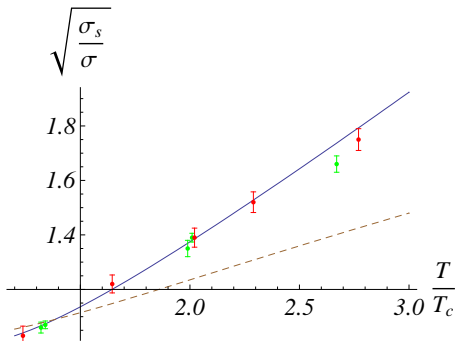}
\caption{\small{Square root of the spatial tension versus temperature. The blue curve corresponds to 
\eqref{stension}. The dots with error bars are from lattice simulations. The green dots are for $N=2$ \cite{bali}, while the red dots are for 
$N=3$ \cite{karsch}. The dashed line represents the result of \cite{agasian}. }}
\end{center}
\end{figure}
We see that our model is in a remarkably good agreement with the available results obtained in lattice 
simulations. In addition, for $T\lesssim 1.6\,\T$ the result is also very close to that obtained analytically in \cite{agasian}.

One point deserves mention here. In Fig.4, we have normalized the spatial tensions at $T=0$ for $N=2$ and 
at $T=1.062560\,\T$ for $N=3$.\footnote{We thank J. Engels for providing us with the numerical values for the points shown in Fig.11 of 
\cite{karsch}.} 

As seen form \eqref{stension}, our model predicts that for the temperature range $1.2\,\T\leq T\leq 3\,\T$ the ratio $\sigma_s/\sigma$ is a 
function of $T^2/T_c^2$. It does not explicitly depend on $N$.\footnote{Interestingly, the pressure normalized by the leading $T^4$-term also 
behaves in a similar way \cite{andreev}.}

\section{Concluding Comments}
\renewcommand{\theequation}{3.\arabic{equation}}
\setcounter{equation}{0}
\noindent (i) The results of this paper together with those of \cite{andreev} provide us with a string picture of how the  
$SU(N)$ gauge theory behaves in the temperature range $1.2\,\T\leq T\leq 3\,\T$ that is remarkably consistent with the available 
results of lattice simulations for $N=2,\,3$. This gives more evidence that string theory may be considered as a real alternative 
to the lattice theory in studying strongly coupled gauge theories. Clearly, string theory is now not on a rigorous foundation and a lot of work is 
required to put the last point. 

\noindent (ii) The string model we have developed predicts that for $1.2\,\T\leq T\leq 3\,\T$ the spatial string tension in units of the physical 
tension is universal in the sense that it depends only on the ratio $T/\T$. There is no explicit dependence on $N$. This is supported by 
lattice simulations for $N=2,\,3$. However, there are no lattice data for larger $N$, and so new simulations would be welcome.

\noindent (iii) The expectation value of a Wilson loop $C$ is generally\footnote{There are many subtleties in \eqref{Wilson}. For some discussion, 
see \cite{witten}.}

\begin{equation}\label{Wilson}
\langle\,W(C)\,\rangle=\int_{\cal D}d\mu\,\ep^{-S(D)}
\,,
\end{equation}
where ${\cal D}$ is the space of string world-sheets obeying boundary conditions and $d\mu$ is a measure of the world-sheet path integral. At 
classical (string) level the integral is evaluated by setting $D$ to the surface of the smallest area. In the problem at hand $S$ is a function of the 
two variable $r$ and $\theta$. Since in the real world there is no dependence on the angle defined in the internal space, one has to get rid 
of it. It seems natural to suggest that at the end of the day \eqref{Wilson} includes the integration over $\theta$ with a covariant measure 
$\sin^4\theta$. If so,  then the expectation value is 

\begin{equation}\label{Wilson1}
\langle\,W(C)\,\rangle=\int_0^\pi d\theta\sin^4\theta \,\ep^{-S(r,\theta)}
\,.
\end{equation}
There is no reason to expect the answer in the form of the exponent needed to extract the potential. Fortunately, there are exceptions and 
the solution, which describes the large distance behavior of the string in the geometry \eqref{metric}, is one of those exceptions. The situation is 
opposite for short distances where the solution does depend on $\theta$.\footnote{See, e.g., \cite{malda, dorn-otto, 
dorn-pershin,sfetos, dorn}.} This makes it difficult to extract the Coulomb term.

Finally, let us note that an alternative recently proposed in \cite{dorn} is to average the "observables'' extracted from $S(r,\theta)$ over $\theta$.

\vspace{.25cm}

{\bf Acknowledgments}

\vspace{.25cm}
\noindent We thank V.I. Zakharov and P. Weisz for stimulating discussions, and J. Engels for providing us with numerical results from lattice
simulations. We also thank H. Dorn for comments and G. Duplancic for help with Mathematica. This work was supported in part
by DFG and Russian Basic Research Foundation Grant 05-02-16486.


\end{document}